\documentclass[12pt]{article}
\usepackage{latexsym,epsfig,graphicx,amsmath,amssymb,amscd,undertilde,multirow,chicago,psfrag,paralist,dsfont,mathtools}
\usepackage{bm}
\usepackage[titletoc]{appendix}
\usepackage{color}
\newcommand{\betavec}{{\boldsymbol{\beta}}}

\newcommand{\D}{\mathcal{D}}

\newcommand{\psivec}{\boldsymbol{\psi}}

\newcommand{\ALE}{\textrm{ALE}}

\newcommand{\omegavec}{\boldsymbol{\omega}}

\newcommand{\tauvec}{\boldsymbol{\tau}}

\newcommand{\Xvec}{\boldsymbol{X}}
\newcommand{\xvec}{\boldsymbol{x}}

\newcommand{\zerovec}{{\boldsymbol{0}}}

\newcommand{\degmodel}{\textrm{DI-VS}}
\newcommand{\nosel}{\textrm{DI-NVS}}
\newcommand{\linear}{\textrm{DI-VSL}}
\newcommand{\hi}{\textrm{DI-KSL}}

\newcommand{\TER}{\textrm{TER}}
\newcommand{\FNR}{\textrm{FNR}}
\newcommand{\FPR}{\textrm{FPR}}

\begin{document}
%%%%%%%%%%%%%%%%%%%%%%%%%%%%%%%%%%%%%%%%%%%%%%%%%%%%%%%%%%%%%%%%%%%%%%%%%%%%%%%%%%%%%%%%%%%%%%%%%%%%%%%%%%%%%%%%%
\title{Building Degradation Index with Variable Selection for Multivariate Sensory Data}

\author{Yueyao Wang$^1$, I-Chen Lee$^2$, Yili Hong$^1$ and Xinwei Deng$^1$ \\
{\small $^1$Department of Statistics, Virginia Tech, Blacksburg, VA 24061}\\
{\small $^2$Department of Statistics, National Cheng Kung University, Tainan, Taiwan 70101}
}

%\date{\today}
%\author{}
\date{}

\maketitle

\begin{abstract}
The modeling and analysis of degradation data have been an active research area in reliability and system health management. As the senor technology advances, multivariate sensory data are commonly collected for the underlying degradation process. However, most existing research on degradation modeling requires a univariate degradation index to be provided. Thus, constructing a degradation index for multivariate sensory data is a fundamental step in degradation modeling. In this paper, we propose a novel degradation index building method for multivariate sensory data. Based on an additive nonlinear model with variable selection, the proposed method can automatically select the most informative sensor signals to be used in the degradation index. The penalized likelihood method with adaptive group penalty is developed for parameter estimation. We demonstrate that the proposed method outperforms existing methods via both simulation studies and analyses of the NASA jet engine sensor data.
	
\textbf{Key Words}: Adaptive LASSO; General Path Model; Prognostics; Sensor Selection; Splines; System Health Monitoring.
\end{abstract}

%\newpage
%\tableofcontents

\newpage
%%%%%%%%%%%%%%%%%%%%%%%%%%%%%%%%%%%%%%%%%%%%%%%%%%%%%%%%%%%%%%%%%%%%%%%%%%%%%%%%%%%%%%%%%%%%%%%%%%%%%%%%%%%%%%%%
\section{Introduction}
%%%%%%%%%%%%%%%%%%%%%%%%%%%%%%%%%%%%%%%%%%%%%%%%%%%%%%%%%%%%%%%%%%%%%%%%%%%%%%%%%%%%%%%%%%%%%%%%%%%%%%%%%%%%%%%%%
\subsection{Background}\label{sec:background}
%%%%%%%%%%%%%%%%%%%%%%%%%%%%%%%%%%%%%%%%%%%%%%%%%%%%%%%%%%%%%%%%%%%%%%%%%%%%%%%%%%%%%%%%%%%%%%%%%%%%%%%%%%%%%%%%%
Degradation data have been a widely-used resource of information for reliability and system health assessment (e.g., \citeNP{meeker1998}). There are many examples of products and systems that provide degradation data. Such as the loss of light output from a light-emitting diode (LED) array, the power output decrease of photovoltaic (PV) arrays, and the vibration from a worn bearing in a wind turbine. The data type is typically a repeated measurement of the degradation index (e.g., the loss of light output from an LED array). Usually, a degradation index has a monotone increasing (decreasing) trend. \citeN{lu1993} is among one of those first that used degradation measurements to assess reliability information, in which the general path model (GPM) is introduced. In the typically modeling framework of the GPM, let $\D(t)$ be the actual degradation path. A soft failure occurs when the degradation level $\D(t)$ reaches a predefined failure threshold. The failure-time random variable is the collection of the first crossing times for all the units in the population. The availability of the degradation measurement is critical for the GPM modeling framework.

The stochastic process model (SPM) framework is also popular in the degradation literature. In the SPM framework, the distribution of the degradation incremental levels is modeled by a Gaussian distribution or other distributions such as the gamma or inverse Gaussian distribution. In other words, the SPM also directly models the degradation measurements. Thus, most existing research on those traditional degradation data modeling assumes that the degradation index for a product or system is well defined and can be measured over time.

Different from traditional degradation data, modern sensor technology allows one to collect multi-channel sensor data that are related to an underlying degradation process. We will discuss one example in detail in Section~\ref{sec:motivating.example}. In such multi-channel sensor data, any single channel may not be sufficient to represent the underlying degradation process. Without a degradation index, most existing methods will not be applicable in the analysis of such sensor signal data. Thus, building a degradation index is an important step in utilizing the sensor data in degradation analysis. Besides, not all recorded sensor signals affect the underlying degradation process. Thus, variable selection is also desirable. This paper aims to develop a flexible method for constructing degradation index from multivariate degradation signals with automatic variable selection.

%%%%%%%%%%%%%%%%%%%%%%%%%%%%%%%%%%%%%%%%%%%%%%%%%%%%%%%%%%%%%%%%%%%%%%%%%%%%%%%%%%%%%%%%%%%%%%%%%%%%%%%%%%%%%%%%%
\subsection{The Motivating Application}\label{sec:motivating.example}
%%%%%%%%%%%%%%%%%%%%%%%%%%%%%%%%%%%%%%%%%%%%%%%%%%%%%%%%%%%%%%%%%%%%%%%%%%%%%%%%%%%%%%%%%%%%%%%%%%%%%%%%%%%%%%%%%

\begin{figure}
	\begin{center}
		\begin{tabular}{cc}
			\includegraphics[width = 0.48\textwidth]{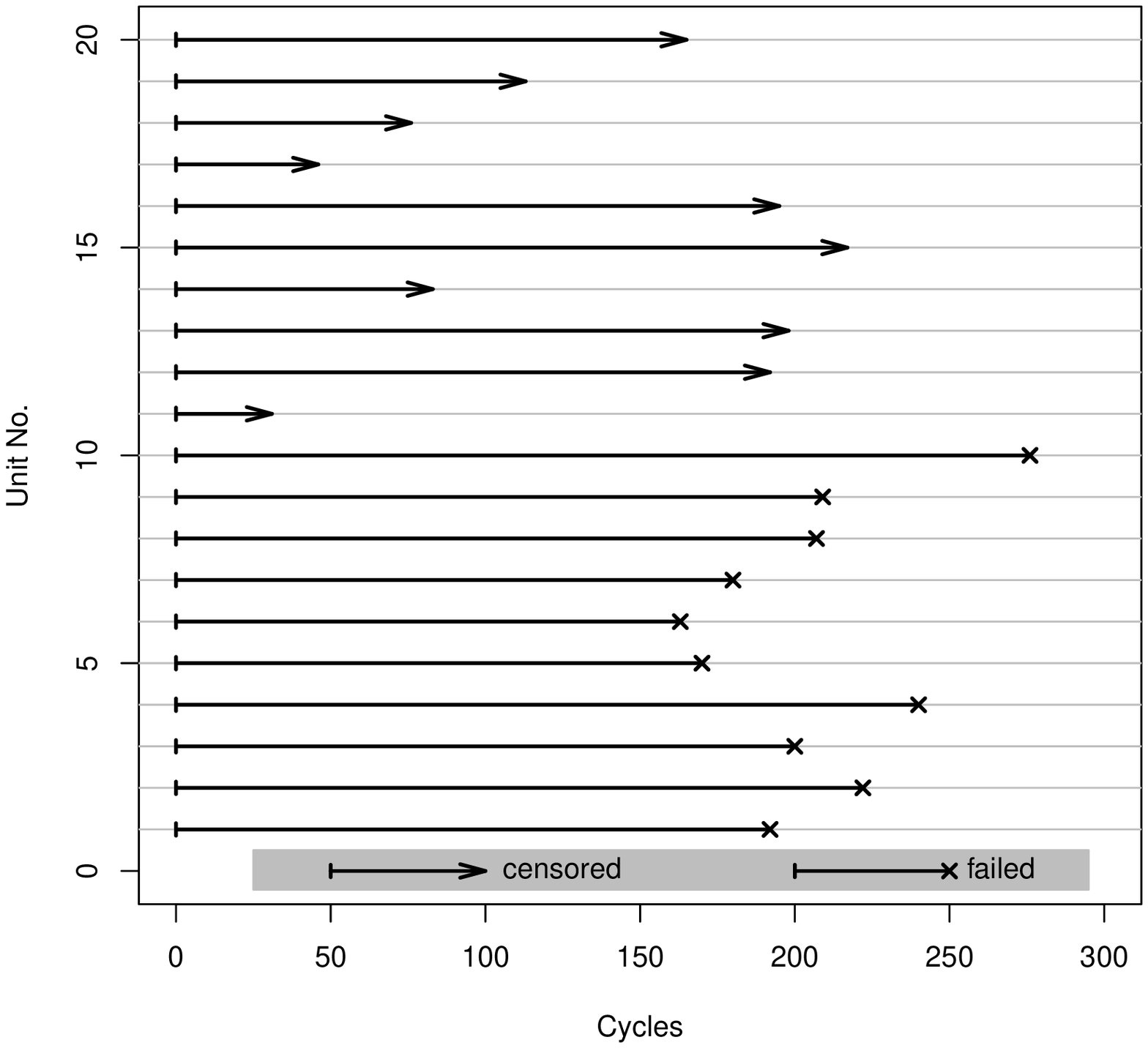}&
			\includegraphics[width = 0.48\textwidth]{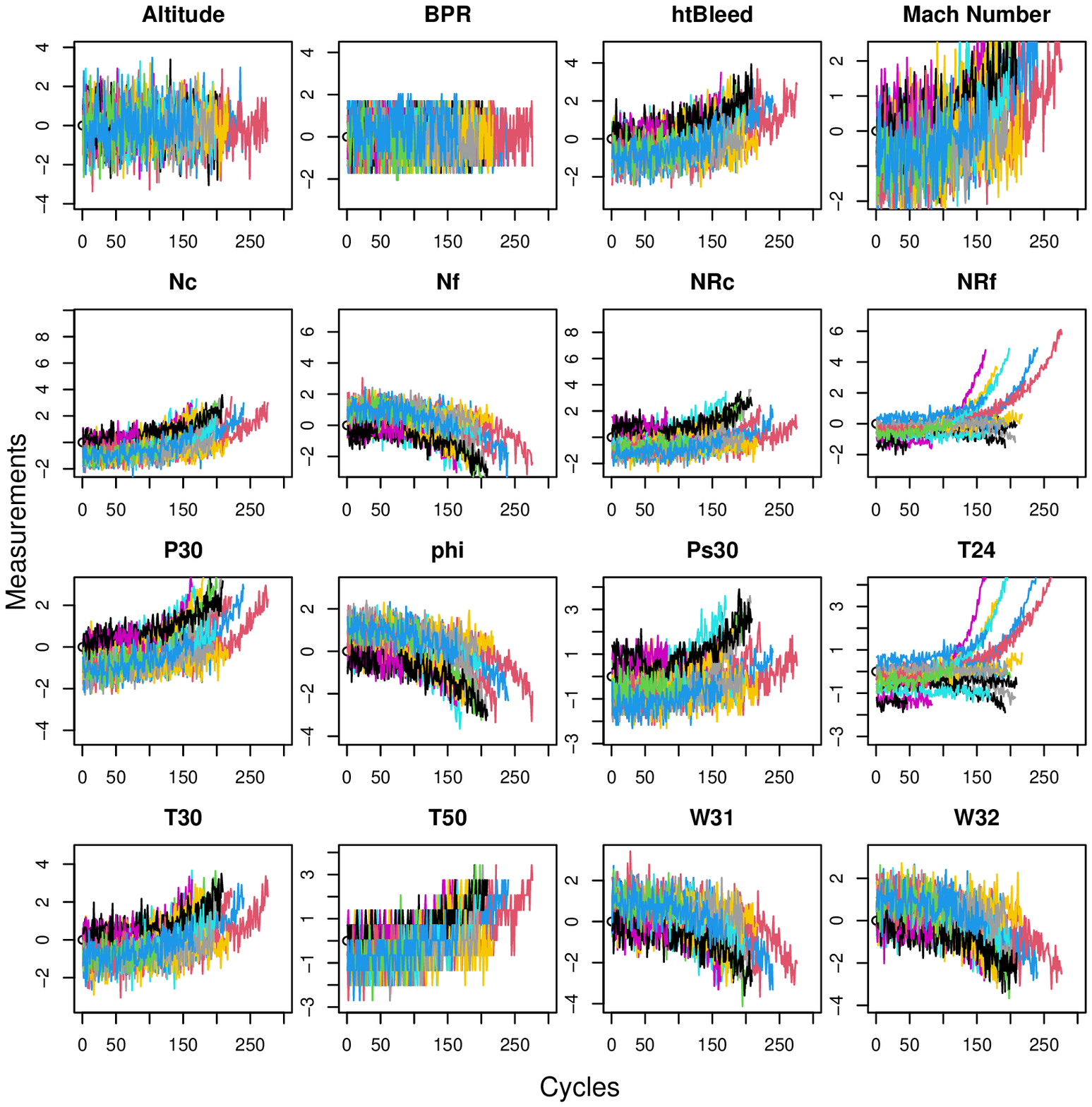}\\
			(a) Time to Event & (b) Multi-channel Signals\\
		\end{tabular}
	\end{center}
	\caption{Plot of a subset of the time to event (a), and multi-channel signals of the 20 units presented by different colors (b).}\label{fig:demo.example}
\end{figure}

The motivating data are the multi-channel sensor data from the jet engine data. The details about the dataset are available in \citeN{SaxenaGoebel2008}. Here we provide a brief introduction. For illustration, we use a subset that contains 200 units, which includes 100 failures and 100 surviving units. The failure-time data give the cycles to failure for failed units and time in service for surviving units. The multi-channel sensor data give time-varying signals with cycles for all 200 units. The dataset includes 21 sensor outputs that measure the system's physical and functional conditions. There are 8 sensors that record temperature and pressure at the fan inlet and different outlets and 8 sensors that capture various fan speeds and coolant bleed in the simulation model. Besides, the measurements of pressure ratio, fuel flow ratio, bypass ratio, burner fuel-air ratio, and bleed enthalpy are also provided in the dataset. In the jet engine simulation data, there are 16 multi-channel sensors after removing those with constant signals. For demonstration, Figure \ref{fig:demo.example} visualizes the cycles to failure for 20 units with their corresponding 16 multi-channel signals.

Although the signal from each channel could be related to the underlying degradation process, none of the sole sensors can be used to represent the degradation process. Thus, it is of interest to build a degradation index from multivariate degradation signals. However, it is typically complicated that how each sensor signal reflects the overall degradation. In such a case, a linear form for the effect of each sensor signal may not be adequate, which motivates us to consider nonlinear functional forms of the individual sensor signal. In most cases, not all sensors are useful in representing the underlying degradation process. Thus, it is important to automatically select more useful sensors to build the degradation index.

On the other hand, censored data are quite common for in the jet engine data and other applications. Thus it is ideal to use both exact failure and censored time data in the training of the model. In addition, the risk of being false positive and false negative are quite different when considering the prediction accuracy, especially for failures of important systems such as jet engines. In this paper, we introduce an asymmetric loss function during the training of the degradation index. Based on those considerations, the jet engine data motivate us to develop a flexible method for construction degradation with automatic variable selection.

%%%%%%%%%%%%%%%%%%%%%%%%%%%%%%%%%%%%%%%%%%%%%%%%%%%%%%%%%%%%%%%%%%%%%%%%%%%%%%%%%%%%%%%%%%%%%%%%%%%%%%%%%%%%%%%%%
\subsection{Literature Review and Contribution of This Work}
%%%%%%%%%%%%%%%%%%%%%%%%%%%%%%%%%%%%%%%%%%%%%%%%%%%%%%%%%%%%%%%%%%%%%%%%%%%%%%%%%%%%%%%%%%%%%%%%%%%%%%%%%%%%%%%%%

Regarding GPM for degradation data, \citeN{nelson1990} introduced constant rate and random coefficient models for accelerated degradation testing data. \citeN{meeker1998} presented a linear path model to obtain the failure-time distribution. \shortciteN{Xieetal2018} proposed a semi-parametric model for accelerated destructive degradation testing. These models assumed an additive relationship between the actual degradation path and a random error term, which together account for the randomness of the observed degradation data. For the SPM for degradation data,  \citeN{Whitmore1995} modeled the degradation process as a Wiener diffusion process with a normal measurement error and the measurements are not assumed to be monotonically increasing. \citeN{ParkPadgett2005} modeled the accelerated degradation for failure based on geometric Brownian motion or gamma process. \citeN{WangXu2010} and \citeN{YeChen2014} proposed models for degradation data based on the inverse-Gaussian process, which ensures the monotone degradation path. In our model, we ensure the monotonic degradation index through the model specification. Regarding sensor data, \shortciteN{HongZhangMeeker2018} and \citeN{MeekerHong2014} outlined some opportunities in using sensor data in reliability modeling and analysis.

In the area of degradation index building, \shortciteN{liu2013} proposed a data-level fusion model for developing composite health indices for degradation modeling and prognostic analysis. \citeN{song2018statistical} combined multiple sensor signals to better characterize the degradation process. \shortciteN{kim2019generic} proposed a latent linear model that constructs a health index via multiple sensors and selects informative sensors. \shortciteN{FangPaynabarGebraeel2017} studied a multi-stream sensor fusion-based prognostics model for systems with a single failure mode. \shortciteN{Chehadeetal2018} considered a data-level fusion approach for degradation index building under multiple failure modes. Sometimes, the linear relationship is not sufficient to capture the system degradation. Thus, this study aims to provide the flexibility to account for the nonlinear relationship between sensors and degradation path.

To construct a meaningful degradation index, it is quite critical to efficiently select useful signals. Regarding variable selections, \citeN{Tibshirani1996} studied the least absolute shrinkage and selection operator (LASSO) penalty for regression-type problems. Later, \citeN{Zou2006} developed the adaptive LASSO to ensure the variable selection consistency. Meanwhile, \citeN{yuan2006} considered the group LASSO for efficient variable selections with meaningful interpretation.
Under the context of building an effective degradation index, we focus on the adaptive group LASSO to select the important sensors.

Specifically, we propose a novel framework based on the cumulative exposure model to build the degradation index with multivariate sensors. To enable sufficient flexibility on the nonlinear relationship between sensors and degradation path, the non-negative spline basis functions are used to describe the contribution of each sensor signal in the cumulative exposure. Thus, we adopt adaptive group LASSO to automatically select the most informative sensors related to the degradation process.

There are several highlights in the proposed framework. The proposed framework is able to include censored failure time information to train the model, which can preserve the information provided by the data. Besides, the proposed model automatically guarantees the monotonicity of the degradation index. We introduce asymmetric loss function and practical failure threshold in the model to improve model prediction accuracy. We also consider nonlinear functions in the model to capture arbitrary forms of sensors impact to the degradation process, which allows great flexibility of the model. With this framework, we are able to build one well-defined degradation index with multi-sensory data so that most existing models for degradation data can be applied.

%%%%%%%%%%%%%%%%%%%%%%%%%%%%%%%%%%%%%%%%%%%%%%%%%%%%%%%%%%%%%%%%%%%%%%%%%%%%%%%%%%%%%%%%%%%%%%%%%%%%%%%%%%%%%%%%%
\subsection{Overview}
%%%%%%%%%%%%%%%%%%%%%%%%%%%%%%%%%%%%%%%%%%%%%%%%%%%%%%%%%%%%%%%%%%%%%%%%%%%%%%%%%%%%%%%%%%%%%%%%%%%%%%%%%%%%%%%%%
The rest of this paper is organized as follows. Section~\ref{sec:building.deg.idx} introduces the framework for degradation index building based on time-to-event data with multivariate degradation signals. Section~\ref{sec:par.est} presents the details for parameter estimation with variable selection. Section~\ref{sec:simulation.study} uses simulation to study the performance of the proposed methods for building degradation index. The motivating example is revisited and is used to illustrate the developed method in Section~\ref{sec:application}. Section~\ref{sec:conclusion} contains conclusions and discusses some potential areas for future research.

%%%%%%%%%%%%%%%%%%%%%%%%%%%%%%%%%%%%%%%%%%%%%%%%%%%%%%%%%%%%%%%%%%%%%%%%%%%%%%%%%%%%%%%%%%%%%%%%%%%%%%%%%%%%%%%%%
\section{Building Degradation Index}\label{sec:building.deg.idx}
%%%%%%%%%%%%%%%%%%%%%%%%%%%%%%%%%%%%%%%%%%%%%%%%%%%%%%%%%%%%%%%%%%%%%%%%%%%%%%%%%%%%%%%%%%
\subsection{Degradation Index}
%%%%%%%%%%%%%%%%%%%%%%%%%%%%%%%%%%%%%%%%%%%%%%%%%%%%%%%%%%%%%%%%%%%%%%%%%%%%%%%%%%%%%%%%%%
Consider sensor data with $p$ degradation signals. Let $\xvec(t)=\{[x_{1}(s),\cdots, x_{p}(s)]': 0\leq s\leq t\}$ be the collection of information for the $p$ signals from a unit, where $x_{j}(s)$ is the $j$th the dynamic covariate information at time $s$, $j = 1, \ldots, p$. We use the cumulative exposure model (aka, the cumulative damage model) to construct the degradation index (e.g., see \citeNP{HongMeeker2013}). The cumulative exposure $u(t)$ for the covariate history $\xvec(t)$ is defined as,
\begin{equation} \label{eq:u(t)}
u(t)=\int_{0}^t{h\left\{\sum_{j = 1}^p f_j[x_j(s);\betavec_j]\right\}}ds,
\end{equation}
where $f_j[x_j(t);\betavec_j]$ represents the effect function of the signal $x_j(t)$ on the degradation index and $\betavec_j$ are parameters that represent the influence of the covariate on the cumulative exposure. Here, $h(z)$ maps the effect to a positive exposure, and the integral is from $0$ to $t$, which guarantees that $u(t)$ is monotonically increasing. In this paper, we use the nonlinear transformation $h(z) = \log[1+\exp(z)]/\log(2)$,
which transforms the input $z \in \left(-\infty, \infty\right)$ to an output that takes value $h(z) \in \left(0, \infty\right)$.

When modeling the contribution of $j$th signal $f_j(\cdot)$ in $u(t)$, it is desirable to make the function form  flexible enough to capture potential nonlinearity in sensors' effect. Therefore, we use a non-negative spline function, called M-splines (e.g., \citeNP{ramsay1988monotone}). For the $j$th signal, let
\[f_j[x_{j}(t);\betavec_j]=\sum_{k=1}^m\beta_{jk}\gamma_{jk}[x_j(t)], \]
where $\{\gamma_{jk}[x_j(t)]: k = 1,\dots, m\}$ are spline bases of the M-spline of order three with $(m-3)$ interior knots and  $\betavec_j = (\beta_{j1}, \ldots, \beta_{jm})^\prime$ are the coefficients of the bases. Let $\betavec = (\betavec_1^\prime, \ldots, \betavec_p^\prime)^\prime$ be the parameters for all $p$ signals. Figure \ref{fig:LEV.spline}(a) shows the bases of the M-spline of order 3 with 7 interior knots ($m=10$). The magnitude of $\betavec$ can be used to identify which signals are more important for the degradation index $u(t)$.

Note that $u(0)=0$, and $u(t)$ is always monotonically increasing, which are the two properties of $u(t)$ that satisfy the characteristics of a degradation index as introduced in Section~\ref{sec:background}. Thus, in this paper, we propose to use $u(t)$ as a degradation index.

%%%%%%%%%%%%%%%%%%%%%%%%%%%%%%%%%%%%%%%%%%%%%%%%%%%%%%%%%%%%%%%%%%%%%%%%%%%%%%%%%%%%%%%%%%
\subsection{Modeling Time to Failure and Degradation Index}\label{sec:modeling.ttf.di}
\begin{figure}%[!ht]
	\centering
	\begin{tabular}{cc}
		\includegraphics[width=.48\textwidth]{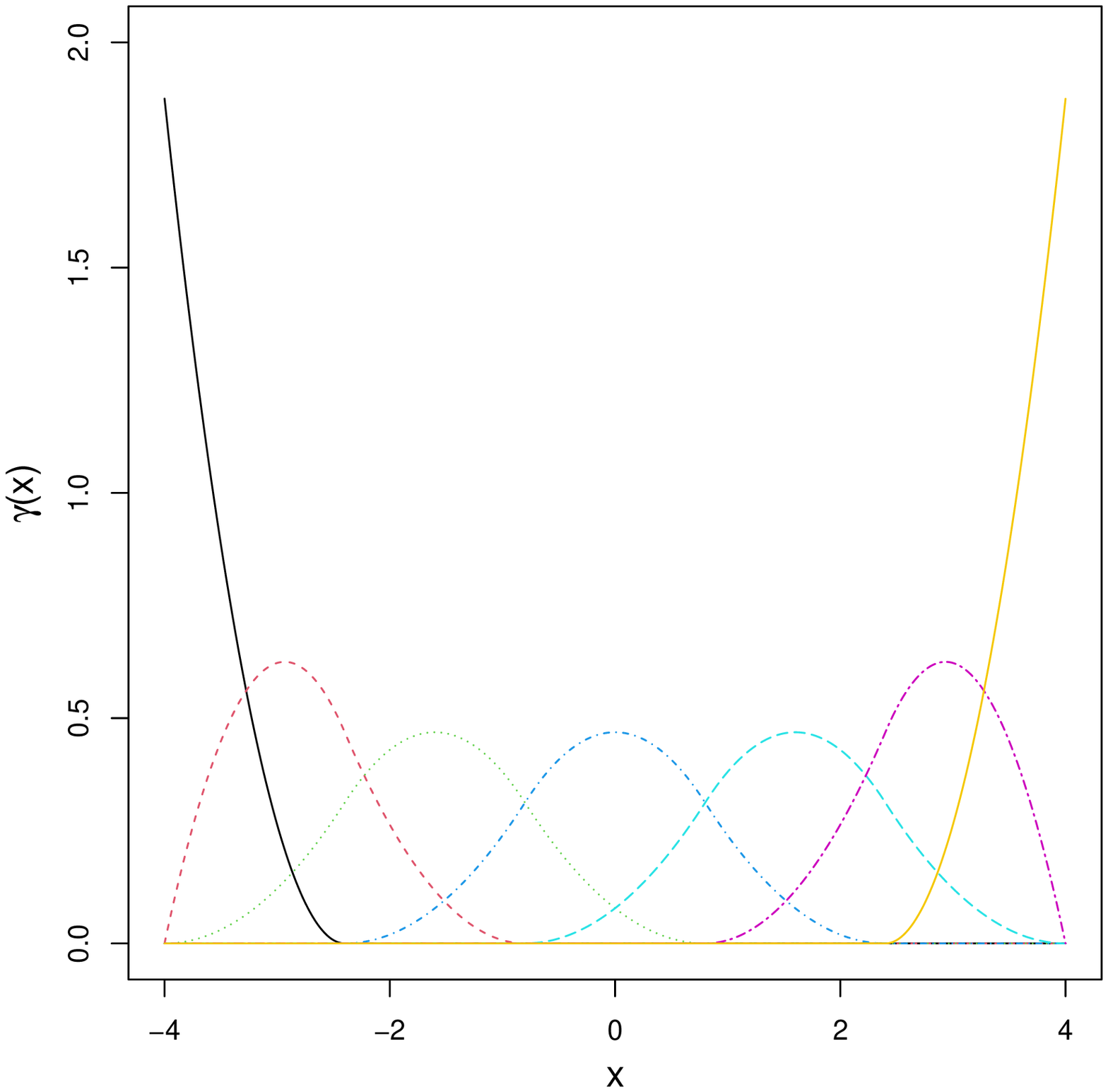} &
		\includegraphics[width=0.48\textwidth]{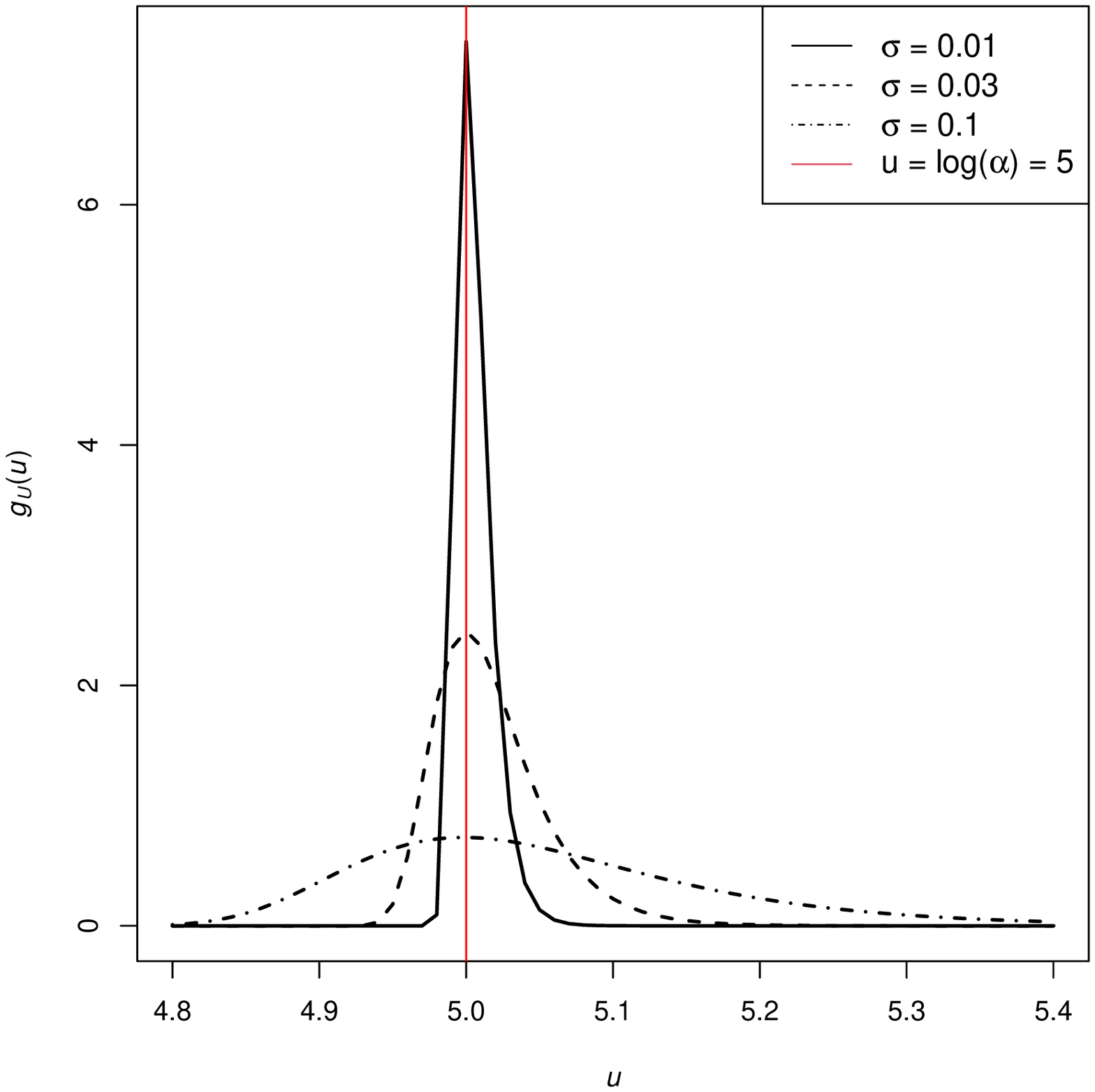} \\
		(a) M-spline Basis & (b) Probability Density Function
	\end{tabular}
	
	\caption{\label{fig:LEV.spline} The plots show the construction of the M-spline of order 3 with 7 interior knots (a), and the pdf of  LEV distributions with $\alpha = \exp(5)$ and various values of $\sigma$ (b).}
\end{figure}

%%%%%%%%%%%%%%%%%%%%%%%%%%%%%%%%%%%%%%%%%%%%%%%%%%%%%%%%%%%%%%%%%%%%%%%%%%%%%%%%%%%%%%%%%%
Based on the cumulative exposure model, a unit fails at time $T$ when the cumulative exposure reaches a random threshold $U$ (e.g., \citeNP{HongMeeker2013}). That is
\begin{align}\label{eqn:U.and.T}
U=u(T),
\end{align}
where the function $u(\cdot)$ is defined in~(\ref{eq:u(t)}). We model $U$ by the largest extreme value (LEV) distribution (e.g., \shortciteNP{de2006extreme}) with the location parameter $\log (\alpha)$ and the scale parameter $\sigma > 0$. The cumulative distribution function (cdf) and the probability density function (pdf) of $U$ can be expressed as
\[
G_U(u;\alpha, \sigma)=\Phi_{\text{LEV}}\left[\frac{\log(u)-\log(\alpha)}{\sigma}\right] \text{ and }
g_U(u;\alpha, \sigma)=\frac{1}{\sigma u}\phi_{\text{LEV}}\left[\frac{\log(u)-\log(\alpha)}{\sigma}\right],
\]
where $\Phi_{\text{LEV}}(x) = \exp[-\exp(-x)]$, and $\phi_{\text{LEV}}(x) = \exp[-x-\exp(-x)]$. As an illustration, Figure~\ref{fig:LEV.spline}(b) shows the pdf of LEV distributions with $\alpha = 5$ and $\sigma = 0.01$, $0.03$, and $0.1$.

The parameter $\alpha$ in the LEV distribution can be used as the \emph{target failure threshold} for the degradation index in \eqref{eq:u(t)}. This is because we want those failed units with their degradation indexes $u(t)$ centered around the failure threshold when they fail. The scale parameter $\sigma$ serves as a measurement of how small the difference between $u(t)$ and the threshold $\alpha$ is if the unit is failed. Because we do not observe $U$, in practice, we use the following threshold rule. That is, if $u(t)\geq \alpha$, we say a unit fails, and if $u(t)<\alpha$, we say the unit is surviving (i.e., the operation status is normal). Through re-scaling, we can map the degradation index to any range that is desirable for the particular application. For example, using $100u(t)/\alpha$, one can map the normal range of the degradation index into $[0, 100]$.

For building the degradation index, a suitable property of the LEV distribution is its skewness to the right as shown in Figure~\ref{fig:LEV.spline}(b). Thus, we allow the larger difference between $u(t)$ and $\alpha$ on the positive side so when making predictions, we can avoid false negative error (i.e., falsely predict failed units as censored). This is desirable because, usually false negative will lead to a higher loss function than false positive, and for important equipment like jet engines, false positive is more tolerable than false negative. %For a smaller value of $\sigma$, the value of $u(t)$ should fall near the threshold $\alpha$ if the unit fails.
Through the relationship between $T$ and $U$ as shown in \eqref{eqn:U.and.T}, the cdf and the pdf of $T$ are,
\[
G_T(t;\betavec)=\Phi_{\text{LEV}}\left\{\frac{\log[u(t)]-\log(\alpha)}{\sigma}\right\},\]
and
\begin{equation}
g_T(t;\betavec)=\frac{u^\prime(t)}{\sigma u(t)}\phi_{\text{LEV}}\left\{\frac{\log[u(t)]-\log(\alpha)}{\sigma}\right\},
\end{equation}
where
\[
u^\prime(t)=h\left(\sum_{j=1}^{p}\sum_{k=1}^m\beta_{jk}\gamma_{jk}[x_j(t)]\right)
\]
is the derivative of $u(t)$.

%%%%%%%%%%%%%%%%%%%%%%%%%%%%%%%%%%%%%%%%%%%%%%%%%%%%%%%%%%%%%%%%%%%%%%%%%%%%%%%%%%%%%%%%%%
%%%%%%%%%%%%%%%%%%%%%%%%%%%%%%%%%%%%%%%%%%%%%%%%%%%%%%%%%%%%%%%%%%%%%%%%%%%%%%%%%%%%%%%%%%%%%%%%%%%%%%%%%%%%%%%%%

\section{Parameter Estimation}\label{sec:par.est}
%%%%%%%%%%%%%%%%%%%%%%%%%%%%%%%%%%%%%%%%%%%%%%%%%%%%%%%%%%%%%%%%%%%%%%%%%%%%%%%%%%%%%%%%%%
\subsection{Log-likelihood with Adaptive Group LASSO Penalty}\label{sec:par.est.likelihood}
%%%%%%%%%%%%%%%%%%%%%%%%%%%%%%%%%%%%%%%%%%%%%%%%%%%%%%%%%%%%%%%%%%%%%%%%%%%%%%%%%%%%%%%%%%
We first introduce some notation for the data. From the sensor data, there are $n$ units with $p$ degradation signals. Three sets of observable data are taken into consideration, which include failure-time data, censoring indicator, and multivariate degradation signals.  Let $t_i$ be the failure time of the $i$th unit and let $\xvec_i(t)=\{[x_{i1}(s),\cdots, x_{ip}(s)]': 0\leq s\leq t\}$ be the collection of $p$ signals for the $i$th unit prior to and at time $t$.  Here $x_{ij}(s)$ is the $j$th observed dynamic covariate information at time $s$, $i=1,\ldots,n$, $j = 1, \ldots, p$, and $0 \leq s \leq t_i$. The observed censoring indicator of the $i$th unit is denoted by $\delta_i$, where $\delta_i$ equals to 1 if the $i$th unit fails and 0 otherwise. Then, the collected information from unit $i$ is denoted by $\left\{ t_i, \delta_i, \xvec_i(t_i) \right\}$, where $i = 1, \ldots, n$, and let $\bm{H} = \left\{\{t_i, \delta_i, \xvec_i(t_i)\}: i=1,\ldots,n\right\}$ be the collection of the data.

To estimate the parameters $\betavec$, the likelihood for the $i$th unit is expressed as,
\begin{align*}
&L_i(\betavec)\\
&= \left(\left[\frac{u'(t_i)}{\sigma u(t_i)}\right]\cdot\phi_{\text{LEV}}\left\{\frac{\log[u(t_i)]-\log(\alpha)}{\sigma}\right\}\right)^{\delta_i}
\times\left(1-\Phi_{\text{LEV}}\left\{\frac{\log[u(t_i)]-\log(\alpha)}{\sigma}\right\}\right)^{1-\delta_i}.
\end{align*}
Note that this is not an actual likelihood because in this case the ``response" $u(t_i)$ is unknown. Instead, the log-likelihood function serves as a loss function for the estimation purpose.
The log-likelihood function for the $i$th unit is
\begin{equation}
\begin{array}{rl} \label{eq:log-l}
l_i(\betavec) =  &  \delta_i \left\{\log[u^\prime(t_i)] - \log (\sigma) - \log[u(t_i)]  - \log\left[\dfrac{\alpha}{u(t_i)}\right]^{1/\sigma} - \left[\dfrac{\alpha}{u(t_i)}\right]^{1/\sigma}  \right\} \\
& + (1-\delta_i) \log \left(1-\exp\left\{-\left[\dfrac{\alpha}{u(t_i)}\right]^{1/\sigma}\right\}\right).
\end{array}
\end{equation}
Hence, the overall log-likelihood function is
\begin{equation} \label{eq:overall_l}
l(\betavec|\bm{H}) = \sum_{i=1}^n l_i(\betavec).
\end{equation}
Then, we can obtain the maximum likelihood (ML) estimates of $\betavec$ by maximizing (\ref{eq:overall_l}).

Note that $\betavec$ are the unknown parameters in the model, and $\alpha$ and $\sigma$ are set as constants because the ``response'' $u(t_i)$ is unknown. From Figure \ref{fig:LEV.spline}(b) and the perspective of the likelihood function, the smaller value of $\sigma$ is, the larger value of the pdf of a failure is. Besides, inside the $\phi_{\text{LEV}}(\cdot)$ and $\Phi_{\text{LEV}}(\cdot)$ calculation, we have the $\log(u) - \log(\alpha) = \log(u/\alpha)$ term. In this case, $\alpha$ and $\betavec$ can take different combinations of values so that $u/\alpha$ keeps the same, which yields the same likelihood. Thus, it is necessary to let $\alpha$ and $\sigma$ be given constants to avoid identifiability issues.

As discussed in Section~\ref{sec:modeling.ttf.di} and from (\ref{eq:log-l}), it can be seen that $u(t_i) \approx \alpha$ if the $i$th unit is a failure, and $u(t_i) < \alpha$ if the $i$th unit is censored. That is, the role of $\alpha$ is to set the target failure threshold. Then, the role of $\sigma$ is to measure how close the difference between $u(t_i)$ and $\alpha$.
So for the value of $\sigma$, ideally, we want $\sigma$ to be a small enough value to allow the degradation index of failed units to end close to the target failure threshold $\alpha$. For the value of $\alpha$, in theory, we can set any value for the target failure threshold $\alpha$. Then $\betavec$ can adjust correspondingly to provide a degradation index between 0 and $\alpha$. In applications, setting $\alpha$ close to the mean of failure times helps the convergence of the estimation algorithm.

Although multiple sensors are available to assess the degradation process, not every sensor collected has a significant contribution. So we integrate variable selection in the model to find out informative sensors. Since there are multiple M-splines basis to represent one sensor, we want to penalize coefficient parameters associated with one sensor simultaneously when that sensor does not contribute. Therefore, we adopt the adaptive group LASSO method \shortcite{huang2010} to conduct variable selection. The adaptive group LASSO approach penalizes parameters in the same group simultaneously. In our model, the parameters in M-splines for the same variable are treated to be in the same group. That is, $\betavec_i$ and $\betavec_j$ are in different groups for any $i \neq j$. From the perspective of variable selection, the $j$th sensor variable has no effect on $u(t)$ if all elements in $\betavec_j$ are significantly small.
The adaptive group LASSO considers the penalties on different grouped parameters have different effects. Let $\omega_j$ be a given weight of the penalty for the $j$th variable, where $\omega_j \geq 0$ and $j=1, \ldots, p$. Then, the penalized negative log-likelihood function is
\begin{equation} \label{eq:obj_AdGL}
\mathcal{L}(\betavec; \lambda) = -l(\betavec|\bm{H}) + \lambda \sum_{j=1}^p \omega_{j} ||\betavec_j||_2,
\end{equation}
where $\lambda \ge 0$ is a tuning parameter and $||\betavec_j||_2 = \sqrt{\sum_{k=1}^m \beta_{jk}^2}$ is the $\textit{L}_2$ norm of the vector $\betavec_j$. Typically, the weights are given by setting
\begin{equation} \label{eq:wts}
\omega_j = \left\{ \begin{array}{cl}
||\widetilde{\betavec}_j||_2^{-\gamma} & \text{ if }||\widetilde{\betavec}_j||_2 > 0  \\
\infty &  \text{ if }||\widetilde{\betavec}_j||_2 = 0,
\end{array}
\right.
\end{equation}
where $\widetilde{\betavec}_j$ is an estimate of $\betavec_j$ and $\gamma \ge 0$ is a hyper-parameter. Here we follow the practice in \shortciteN{huang2010} and define $\infty \cdot 0 = 0$. That means the model does not select sensor $j$ if its coefficient estimates $\textit{L}_2$ norm is zero (i.e., $||\widetilde{\betavec}_j||_2 = 0$).

%%%%%%%%%%%%%%%%%%%%%%%%%%%%%%%%%%%%%%
\subsection{Optimization of Objective Function}\label{sec:optim}
The optimization of \eqref{eq:obj_AdGL} is challenging. Here we discuss some strategies used in the optimization of the objective functions.  To optimize the objective function, we use the Nelder-Mead algorithm in the R package \textit{nloptr} \cite{nloptr}.

One thing to notice is the influence of $\sigma$ value in the optimization procedure. If $\sigma$ is prefixed at a small value (e.g., 0.01) at the beginning of the optimization, the algorithm could be easily trapped at local optima. As shown in Figure~\ref{fig:LEV.spline}(b), when $\sigma$ is small, the pdf of LEV is highly concentrated around the location parameter $\log(\alpha)$. That means a unit with $\log[u(t_i)]$ at the event time that is close to the threshold $\log(\alpha)$ has a high probability. While a unit which degradation index at the event time is far away from the threshold has almost zero probability, thus, its contribution to the likelihood function is small. During the optimization process, with a small $\sigma$, it is possible that the $\betavec$ is updated to an estimation that some units' degradation paths get almost 0 probability. So the contribution of these units to the objective function is neglected in the following updates of the $\betavec$ estimation. Only units with $\log[u(t_i)]$ that are relatively close to $\log(\alpha)$ have the chance to further move close to the threshold.

One approach to avoid the local optima is to set a larger value for $\sigma$ at the beginning of the optimization process, and then decrease it gradually to the prefixed lower bound. By setting $\sigma$ to a large value, say $\sigma = 1$, the information of all units is equally treated in the objective function, regardless of the distance between the value of $\log[u(t_i)]$ and the location parameter. After a certain number of iterations, the unit's $\log[u(t_i)]$ moves closer to $\log(\alpha)$, then we can decrease the value of $\sigma$ by a small amount and update $\betavec$ estimation. Repeating this step until $\sigma$ decreased to the fixed constant can help to avoid the local optima problem. In this paper, instead of manually determining a sequence of $\sigma$
to decrease, we add it to the optimization parameters.

Let $\sigma_l$ be the prefixed lower bound of $\sigma$. We consider the transformation $$\log(\sigma^{\ast}) = \log(\sigma - \sigma_l).$$ Thus, $\sigma = \exp[\log(\sigma^{\ast})] + \sigma_l$. Then we optimize $\log(\sigma^{\ast})$ and $\betavec$ simultaneously. This transformation can always impose a lower bound for the estimation of $\sigma$. Although we include $\log(\sigma^{\ast})$ in the parameter estimation, the purpose is not to obtain an estimation of $\log(\sigma^{\ast})$. The reason is that $\sigma$ is not identifiable and it always becomes smaller to allow a larger likelihood. Via the iterations, it will get to its lower bound eventually. Thus, we include $\sigma$ in the parameter estimation so that it can smoothly decrease and help to avoid the local optima of $\betavec$ estimation.

With a larger value of $\sigma$ in the early stage of the iterations, the benefit of the asymmetric property of the LEV distribution is not evident. We introduce the following remedy to ensure the estimates of $u(t_i)$ are moving towards the right direction during the early stage of the optimization iterations. As discussed above, when building a degradation index, it is desirable that $u(t_i) = \alpha$ if $i$th unit fails and $u(t_i) < \alpha$ if $i$th unit is censored. Thus, we further impose those two constraints on the objective function. That is, we modify the objective function as,
\begin{equation}
\mathcal{M}\left(\betavec, \lambda\right) = \mathcal{L}(\betavec; \lambda) +
\eta \sum_{i=1}^{n}\delta_i\{[ \alpha  - u(t_i)]\}^2+(1-\delta_i)\{[u(t_i) - \alpha ]_{+}\}^2.
\label{eq:final.obj}
\end{equation}
The positive part function is $[u(t_i) - \alpha]_{+} = \max\left(u(t_i) - \alpha,0\right)$ and the penalty $\eta$ is non-negative. Therefore, to encourage the estimated $u(t)$ satisfying the degradation index characteristic (i.e., $u(t_i) = \alpha$ if $i$th unit fails and $u(t_i) < \alpha$ if $i$th unit is censored) as well as perform variable selection, we work with the objective function $\mathcal{M}\left(\betavec, \lambda\right)$ as shown in \eqref{eq:final.obj}.

\subsection{Determining Tuning Parameter}\label{sec:determining.lambda}
In parameter estimation, we need to determine the tuning parameter $\lambda$ in the objective function (\ref{eq:final.obj}). The $k$-folds cross-validation ($k$-CV) approach is used. Because the main goal of our degradation index model is to accurately predict the status of testing units, especially for the failed units, we use both the false negative error rate and total error rate as the criterion to select the tuning parameters. Let $\FNR$ and $\FPR$ be the averaged false negative and positive error rates across the $k$-folds, respectively. Then the averaged total error rate is $\TER = \FNR + \FPR$.  For a sequence of values for $\lambda$, denoted by $\{\lambda_1, \dots, \lambda_q\}$, the corresponding error rates are $\{\TER_1, \dots, \TER_q\}$, and $\{\FNR_1, \dots, \FNR_q\}$. Let $k_f = \arg\min_b \FNR_b$ be the index of tuning parameter that minimizes $\FNR$. %and $r = \arg\min_i \TER_i$.
We want to select the tuning parameter $\lambda$ so that $\FNR$ is minimized, while $\TER$ is kept at a relatively low level. That means we do not want to sacrifice $\FPR$ to achieve the smallest $\FNR$. Therefore, the selected tuning parameter is $\lambda_s$, of which the index $s$ is determined by
\begin{equation}\label{eq:lambda}
s = \begin{cases}
\arg\min_b \FNR_b & \text{if } \TER_{k_f} \leq 0.2, \\
\arg\min_b \TER_b & \text{otherwise.}
\end{cases}
\end{equation}
In this way, when a $\lambda$ minimizes $\FNR$ at the cost of $\FPR$, we will switch to the $\lambda$ that minimizes $\TER$ to achieve a balance between $\FPR$ and $\FNR$.

The parameters $\eta$, $\gamma$, and $\sigma_l$ are treated as hyper-parameters. We do not tune $\eta$ because the role of the $\eta$ penalty term is to help $u(t)$ move towards $\alpha$ at the beginning stage. After $u(t)$ is close to $\alpha$, the effect of asymmetric property of LEV kicks in and the shrinkage of $\sigma$ towards $\sigma_l$ serves the same role. Therefore, we only include $\eta$ penalty term to help the algorithm converge and set a moderate large value for $\eta$. In particular, $\eta$ is set to be 5. We fix the hyper-parameter $\gamma = 2$ in the calculation of the weights in~(\ref{eq:wts}), which is as a common practice. The lower bound of scale parameter $\sigma_l$ is set to be 0.01.

%%%%%%%%%%%%%%%%%%%%%%%%%%%%%%%%%%%%%%%%%%%%%%%%%%%%%%%%%%%%%%%%%%%%%%%%%%%%%%%%%%%
\subsection{Parameter Estimation Procedure}\label{sec:algorithm}
In this section, we describe how to obtain the estimates $\widehat{\betavec}$ based on the training set, using the adaptive group LASSO procedure. With initial estimates $\widetilde{\betavec}_j$, we obtain the weights $\omega_j$, as in~\eqref{eq:wts} for $j=1, \cdots, p$.

We first apply the $k$-CV in Section~\ref{sec:determining.lambda} to obtain the tuning parameter $\lambda_s$ as in \eqref{eq:lambda}. The average of the $\betavec$ estimates from each fold in $k$-CV can also provide a warm starting point for the final estimate of $\betavec$. Similar ideas are also used in literature (e.g., \shortciteNP{mazumder2011sparsenet}). With the best selected $\lambda_s$, the warm starting point for $\betavec$, and the entire training set, we apply the adaptive group LASSO procedure to obtain the final estimates of $\betavec$ in (\ref{eq:final.obj}), which is denoted by $\widehat{\betavec}$. The statistical inference and variable selection results can be obtained based on $\widehat{\betavec}$.

The question remains how to find the initial estimates $\widetilde{\betavec}_j$'s.  \shortciteN{huang2010} suggested that one can use the group LASSO procedure (i.e., without adaptive weights) to find the initial estimates of parameters. For the group LASSO estimates, the objective function in \eqref{eq:final.obj} is simplified by setting $\omega_j = 1, j=1, \cdots, p$. Similarly, we apply the $k$-CV to find the best tuning parameter $\widetilde{\lambda}_s$ for the group LASSO, and use the averaged estimates as the warm starting points for the final estimates of the group LASSO procedure, which are denoted by $\widetilde{\betavec}_j, j=1, \cdots, p$.

For the initial values of the $k$-CV of the group LASSO procedure, we have to use cold starting points as we do not have much information about the parameters at this step. We randomly select the starting points that satisfy some desirable properties in the degradation scenario. For example, the starting points need to allow the minimum of $\log[u(t_i)]$'s of the failed units to be larger than the maximum of $\log[u(t_i)]$'s of the censored units.

%%%%%%%%%%%%%%%%%%%%%%%%%%%%%%%%%%%%%%%%%%%%%%%%%%%%%%%%%%%%%%%%%%%%%%%%%%%%%%%%%%%%%%%%%%%%%%%%%%%%%%%%%%%%%%%%%
\section{Simulation Study}\label{sec:simulation.study}
%%%%%%%%%%%%%%%%%%%%%%%%%%%%%%%%%%%%%%%%%%%%%%%%%%%%%%%%%%%%%%%%%%%%%%%%%%%%%%%%%%%%%%%%%%%%%%%%%%%%%%%%%%%%%%%%%
In order to evaluate the performance of the proposed degradation index building method, we use various simulated datasets to compare our model with the model without variable selection and the model assumes a linear relationship between sensors and degradation index. We want to investigate the average times that our model properly selects variables and the average prediction accuracy (i.e., correctly predicts the status of a unit as a failure or censored one).
%%%%%%%%%%%%%%%%%%%%%%%%%%%%%%%%%%%%%%%%%%%%%%%%%%%%%%%%%%%%%%%%%%%%%%%%%%
%%%%%%%%%%%%%%%%%%%%%%%%%%%%%%%%%%%%%%%
\subsection{Simulation Setup and Procedure}\label{sec:simulation.data}
%%%%%%%%%%%%%%%%%%%%%%%%%%%%%%%%%%%%%%%%%%%%%%%%%%%%%%%%%%%%%%%%%%%%%%%%%%%%%%%%%%%%%%%%%%%%%%%%%%%%%%%%%%%%%%%%%

\begin{figure}
	\centering
	\includegraphics[width=0.65\textwidth]{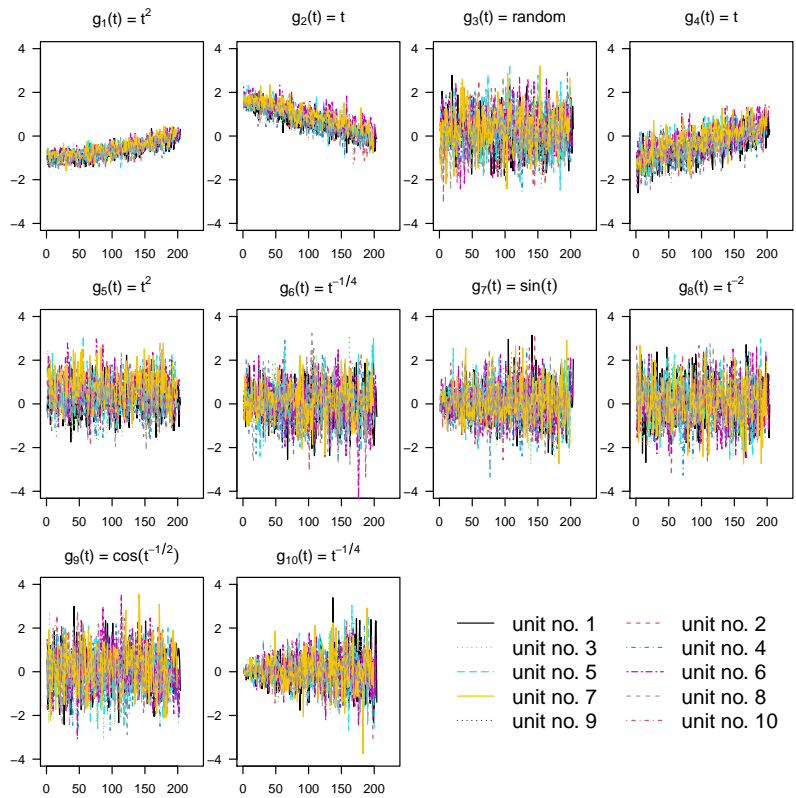}
	\caption{Plot of a subset of units with 10 simulated signals. Each panel represents one signal. The $x$-axis shows the time and the $y$-axis shows the signal value.}\label{fig:sim.signals}
\end{figure}

In this simulation study, we want to generate datasets similar to the jet engine data to demonstrate the performance of our degradation index building framework. To generate the simulation data, we first consider the signal sensors. In the simulation study, we generate 10 sensor signals similar to the jet engine signals within time interval $[0, 350]$. We assume each signal is a function of time with some variations. That is $X_j(t) = g_j(t) + \epsilon_j(t), j = 1, \dots, 10$. The function $g_j(t)$ can take forms such as constant, linear, quadratic, log functions, and the error term $\epsilon_j(t)$ follows a normal or a uniform distribution. Figure~\ref{fig:sim.signals} presents the example of simulated signals for 10 units. The signals can be increasing, decreasing, or randomly fluctuate over time. After obtaining the sensor information, the basis functions of the M-spline with 2 interior knots are constructed based on the signals. So we have 50 coefficient parameters for the simulated data. To test the variable selection capability of our method, we assume that 5 out of 10 signals cause the units to fail, and the rest 5 have no effect on the failure process. Hence, we set the values of parameters of the first 5 signals to have effects on the degradation index. Among the five signals with effect, we assume the second and fourth are linear functions of time, and the first and the fifth are quadratic functions of time, and the third one is assumed to follow a normal distribution. With the parameter coefficients and the simulated signals, we can calculate the simulated $u(t)$. The next step is to generate the failure-time data. The failure-time data are generated by the following steps:

\begin{enumerate}[(a)]
	\item {\bf Generate the censoring time:} Let the time to failure be $T = \min\{C, 350\}$,
	where $C$ follows a Weibull distribution. The shape and scale parameters are determined to ensure the proportion of failed units does not exceed 90\%. The shape and scale parameters can be varied with different coefficient parameters.
	\item {\bf Determine the status of each unit:} Set the failure threshold as $\alpha = \exp(5)$, and the censoring indicator is defined as
	\[ \delta = \left\{ \begin{array}{cl}
	1 &  u(T) \geq \exp(5),\\
	0 & u(T) < \exp(5).
	\end{array}
	\right.
	\]
\end{enumerate}

To test the procedure under different situations, we consider various $n$ and $\betavec$ to control the number of total units and effects degree of covariates. In particular,

\begin{enumerate}
	\item Number of units $n = 50, 100, 150, 200, 250, 300$.
	\item Degree of effects on the degradation process. Because we have 5 splines for each covariate, the corresponding $\betavec_j$ is a vector of length 5 for $j = 1, \ldots, 10$. Assume that the last 5 signals do not affect the degradation process, so $\betavec = \left(\betavec_1', \betavec_2', \betavec_3', \betavec_4',\betavec_5',\zerovec_{25}'\right)'$. We consider the following four scenarios and their corresponding coefficient values are listed in Table~\ref{tbl:betas}. In particular,

	\begin{enumerate}[(A)]
		\item The contribution of effective covariates (i.e., $x_1(t), \dots, x_5(t)$) are on the same magnitude;
		\item The signals with quadratic function forms (i.e., $x_1(t)$ and $x_5(t)$) have larger effects;
		\item The signals with linear function forms (i.e., $x_2(t)$ and $x_4(t)$) have larger effects;
		\item Only the random term (i.e., $x_3(t)$) has larger effect.
	\end{enumerate}
\end{enumerate}

\begin{table}
	\centering
	\caption{The true $\betavec$ under four simulation scenarios.}
	\vspace{2ex}
	\begin{tabular}{r|r|r|r|r|r}
		\hline
		\hline
		& $\betavec_1$ &
		$\betavec_2$ & $\betavec_3$ & $\betavec_4$ & $\betavec_5$  \\
		\hline
		\multirow{5}{*}{Scenario A: $\betavec_a$} & 10.61 & $-$2.49 & $-18.00$ & $-3.07$ & $-$4.86  \\
		& 1.39 & 0.24 & $-3.31$ & $-16.09$ & 11.55  \\
		& 2.75 & $-$4.93 & $-$0.47 & 1.32 & 1.53  \\
		& 4.35 & 0.17 & $-$1.06 & $-$0.13 & 1.97  \\
		& $-$0.38 & $-$16.67 & 2.31 & 7.90 & $-$6.59  \\
		\hline
		\multirow{5}{*}{Scenario B: $\betavec_b$}  & 2.35 & 0.08 & 0.29 & $-$0.12 & $-$2.70  \\
		& 1.76 & $-$0.04 & $-$0.28 & 0.00 & $-$2.09  \\
		& 2.04 & $-$0.18 & 0.30 & 0.14 & $-$2.16  \\
		& 1.59 & $-$0.06 & $-$0.11 & $-$0.08 & $-$2.49  \\
		& 1.74 & $-$0.09 & $-$0.24 & $-$0.04 & $-$1.72  \\
		\hline
		\multirow{5}{*}{Scenario C: $\betavec_c$}  & $-$0.06 & 2.42 & $-$0.42 & $-$2.64 & 0.02  \\
		& 0.16 & 1.44 & 0.48 & $-$1.37 & 0.08  \\
		& $-$0.04 & 2.40 & $-$0.49 & $-$1.59 & 0.10  \\
		& 0.11 & 1.32 & 0.04 & $-$2.54 & 0.07  \\
		& $-$0.09 & 1.79 & 0.47 & $-$1.94 & $-$0.14  \\
		\hline
		\multirow{5}{*}{Scenario D: $\betavec_d$}  & $-$0.32 & $-$0.13 & 3.25 & $-$0.75 & 0.79\\
		& $-$0.46 & 0.24 & 3.05 & $-$0.77 & 0.89  \\
		& 0.02 & $-$0.55 & 3.68 & 0.55 & 0.68  \\
		& $-$0.49 & $-$0.05 & 2.73 & $-$0.18 & 0.63  \\
		& $-$0.47 & 0.28 & 2.99 & 0.78 & 0.66\\
		\hline
		\hline
	\end{tabular}
	\label{tbl:betas}
\end{table}

%%%%%%%%%%%%%%%%%%%%%%%%%%%%%%%%%%%%%%%%%%%%%%%%%%%%%%%%%%%%%%%%%%%%%%%%%%%%%%%%%%%%%%%%%%%%%%%%%%%%%%%%%%%%%%%%%
\subsection{Method Comparisons}\label{sec:comparemodel1}

In order to better understand the performance of the proposed model and procedure, we compare the proposed model with the model without variable selection, and with the model considering only linear relationship for the covariate effect.

For the model without variable selection, $\lambda$ is set to 0 in (\ref{eq:final.obj}). The objective function is as follows,
\begin{equation*} \label{eq:obj_nosel}
\mathcal{M}(\betavec) = -l(\betavec|\bm{H}) + \eta \sum_{i=1}^{n}\delta_i\{[ \alpha  - u(t_i)]\}^2+(1-\delta_i)\{[u(t_i) - \alpha ]_{+}\}^2.
\end{equation*}
If we assume the covariate effect is in a linear form, then degradation index becomes
\begin{equation*} \label{eq:u(t)_linear}
\widetilde{u}(t)=\int_{0}^t h\left\{\sum_{j=1}^{p}x_j(s)\beta_j\right\}ds.
\end{equation*}
We also use adaptive LASSO to perform variable selection and the objective function is the same as \eqref{eq:final.obj} but with $u(t)$ replaced by $\widetilde{u}(t)$.

%%%%%%%%%%%%%%%%%%%%%%%%%%%%%%%%%%%%%%%%%%%%%%%%%%%%%%%%%%%%%%%%%%%%%%%%%%%%%%%%%%%%%%%%%%%%%%%%%%%%%%%%%%%%%%%%%
\subsection{Simulation Results}
%%%%%%%%%%%%%%%%%%%%%%%%%%%%%%%%%%%%%%%%%%%%%%%%%%%%%%%%%%%%%%%%%%%%%%%%%%%%%%%%%%%%%%%%%%%%%%%%%%%%%%%%%%%%%%%%%

We generate simulation data based on the procedure described in Section~\ref{sec:simulation.data} and apply three different models to the simulated data. We denote our proposed model as $\degmodel$, the model without variable selection as $\nosel$, and the model assumes linear sensor effect as $\linear$. In this simulation, the target failure threshold is set to be $\alpha = \exp(5)$. For each sample size $n$ and coefficient parameter vector $\betavec$, we repeat the trial 200 times.

\begin{figure}
	\centering
	\includegraphics[width=0.9\textwidth]{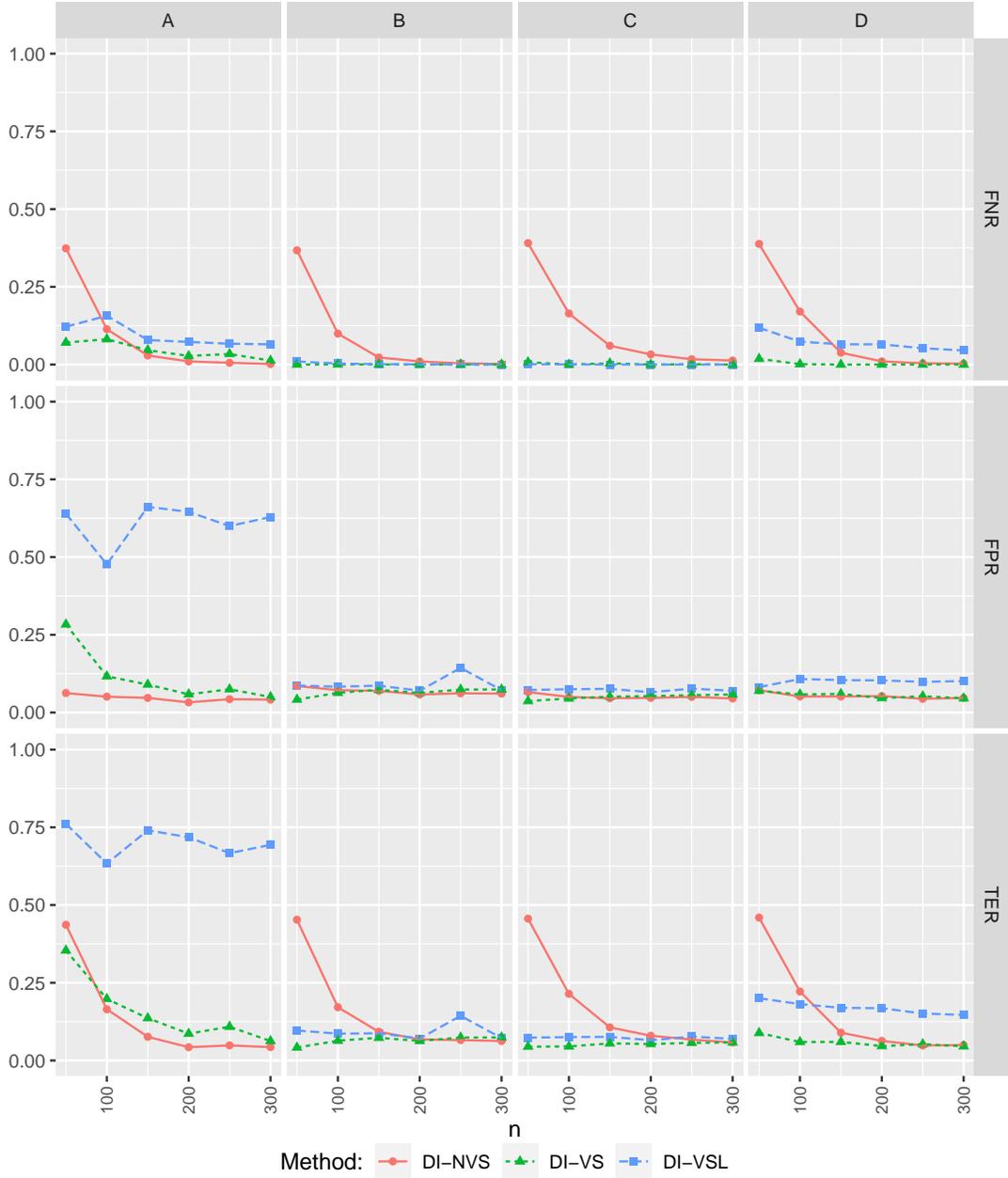}
	\caption{Average FNR, FPR and TER versus number of units ($n$) for different methods and scenarios.}
	\label{fig:sim.error}
\end{figure}

\begin{figure}
	\centering
	\includegraphics[width=0.9\textwidth]{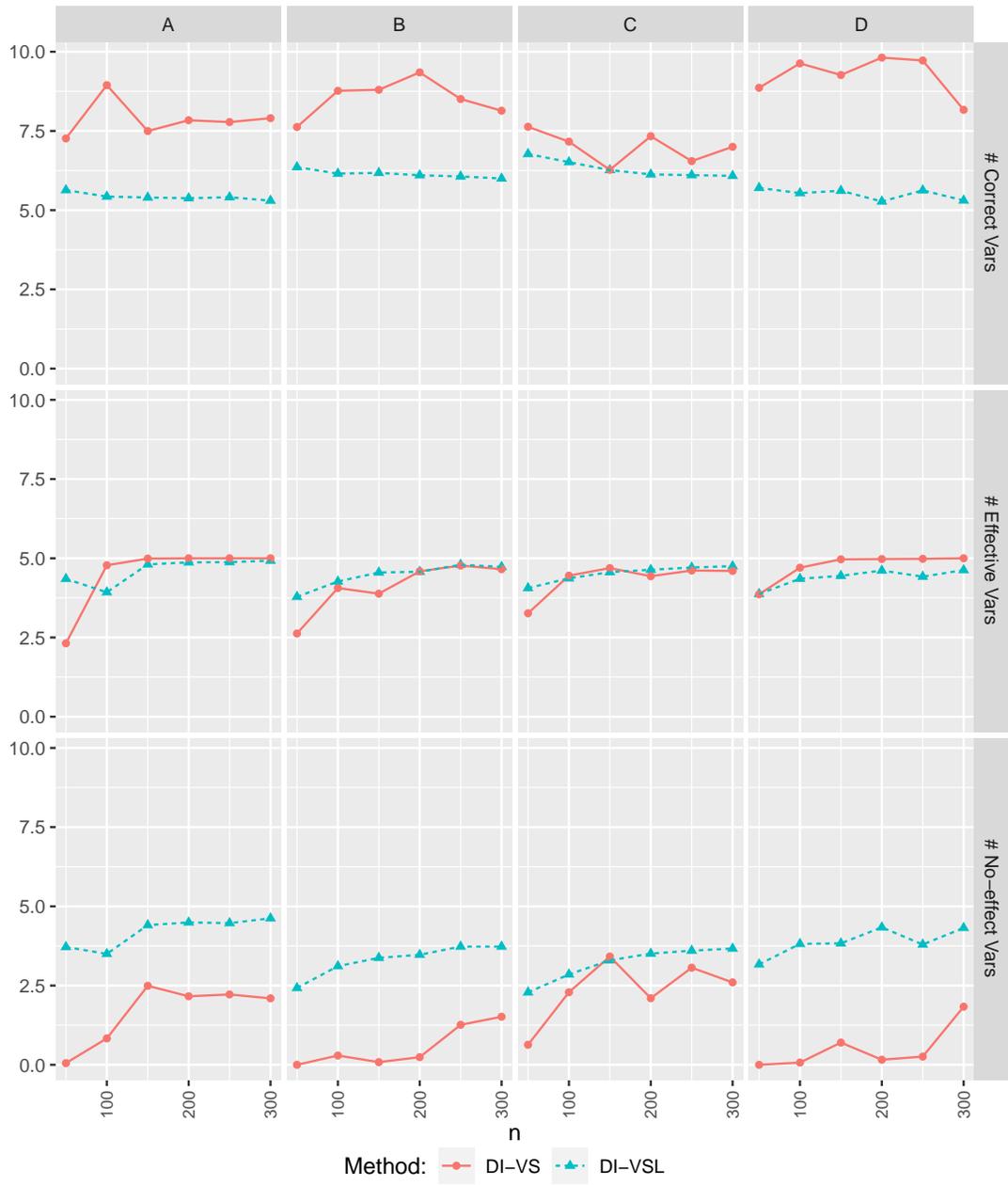}
	\caption{Average number of correctly specified variables, effect variables, and no effect variables identified by the model versus number of units ($n$) for different methods and scenarios. The actual number of effective variables is 5, no-effect variables in the model is 0, and correctly specified variables is 10.}
	\label{fig:sim.var}
\end{figure}

Regrading to the predictions of unit status, although the target failure threshold is $\alpha$, in practice, due to training errors, one can use a threshold that is slightly smaller than $\alpha$, which typically yields better classification results. We call this the  \emph{practical threshold} $\widetilde{\alpha}$. One can choose $\widetilde{\alpha}_p = \exp\left[\log(\alpha) + z_p \sigma\right]$, where $z_p$ is the quantile function of the standard LEV distribution. In the simulation study, the classification uses $\widetilde{\alpha}_{0.01}$ threshold. The simulation results are summarized in Figures~\ref{fig:sim.error} to~\ref{fig:sim.var}.

Figure~\ref{fig:sim.error} shows the average $\FNR$, $\FPR$, and $\TER$ as the number of units increases. Although in general $\FNR$ decreases for all three models, $\degmodel$ has the most consistent performance across various scenarios and the number of units. When the number of units is small (i.e., less than 100), the $\FNR$ of $\nosel$ is the largest among the three models, which shows the benefit to consider variable selection especially when $n$ is relatively small. The $\FNR$ for $\linear$ is large in Scenarios A and D, even when the number of units is large. For $\FPR$, in Scenarios B, C, and D, $\FPR$ does not change a lot with the size increases. In Scenario A, $\nosel$ has the smallest $\FPR$ across the number of units while $\linear$ has the largest. The results show that ignoring nonlinear relationships can lead to larger errors in some scenarios. With respect to the $\TER$, $\degmodel$ has the smallest errors for almost all scenarios.

Regrading the variable selection capability, Figure~\ref{fig:sim.var} shows the number of signals effect and no effect in the model as well as the correctly specified variables. Ideally, we want the model to include all 5 effective variables and zero no-effect variable. The number of correctly specified variables includes the number of effective variables remained in the model and no-effect variables excluded from the model, which should be 10. Compared to $\linear$, $\degmodel$ tends to include more effective signals in the model when the number of units is large in Scenarios A and D. For Scenarios B and C, the number of effect signals remained in the model are close to $\linear$ when the number of units is large. For signals that have no effect on the underlying degradation process, $\degmodel$ can exclude more signals across all four scenarios and various numbers of units. In general, $\degmodel$ has more correctly specified variables across different simulation scenarios compared to $\linear$.

Overall, the simulation results show that our proposed model has better accuracy in predicting a unit status for various scenarios and numbers of units. It can also exclude signals with no effect from the model.

\begin{figure}
	\centering
	\includegraphics[width=0.9\textwidth]{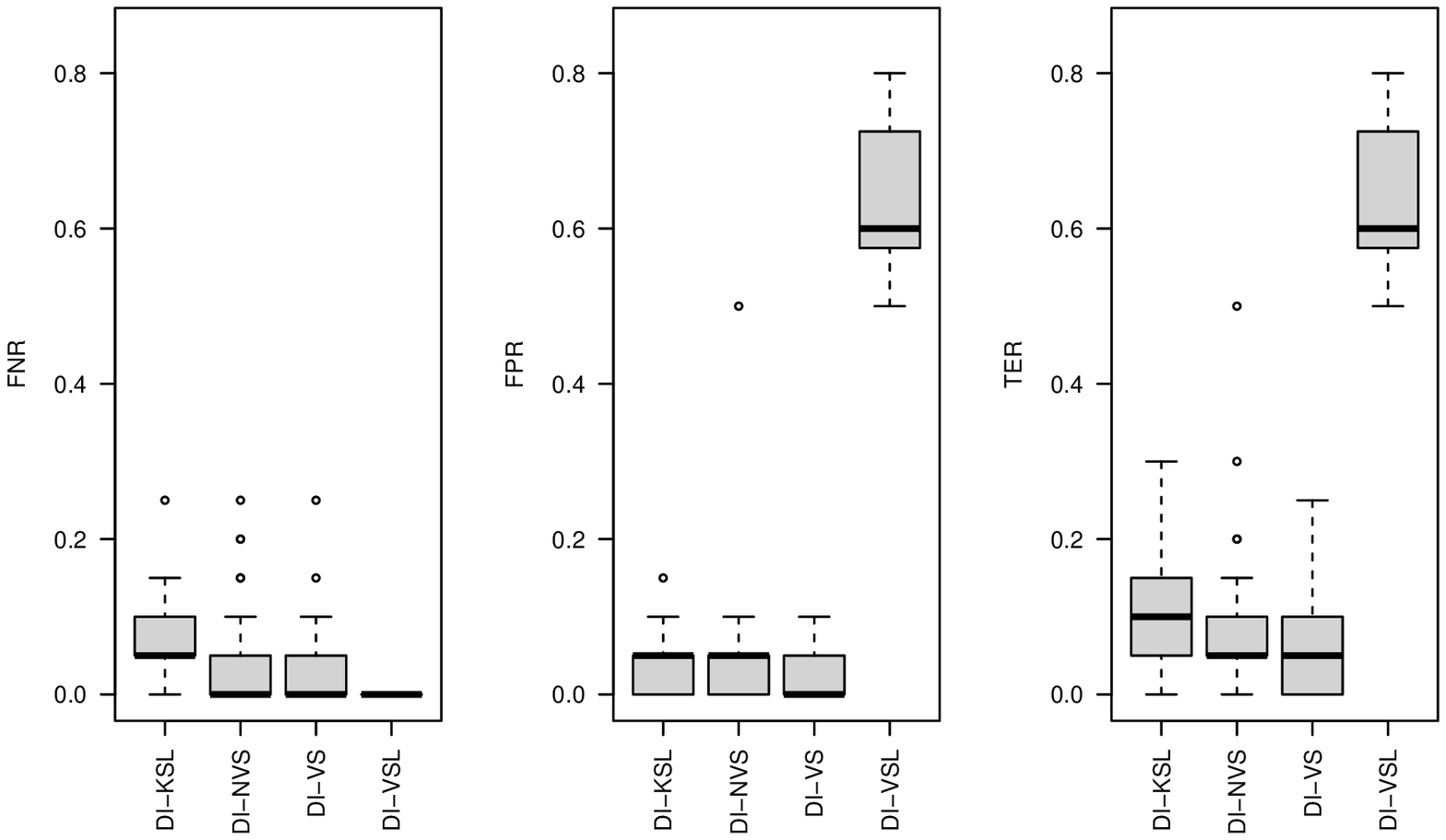}
	\caption{The boxplot of error rates with jet engine data with 50 replicates for different methods with $\widetilde{\alpha}_{0.01}$ practical threshold.}
	\label{fig:real.compare001}
\end{figure}

%%%%%%%%%%%%%%%%%%%%%%%%%%%%%%%%%%%%%%%%%%%%%%%%%%%%%%%%%%%%%%%%%%%%%%%%%%%%%%%%%%%%%%%%%%%%%%%%%%%%%%%%%%%%%%%%%
\section{Application}\label{sec:application}
%%%%%%%%%%%%%%%%%%%%%%%%%%%%%%%%%%%%%%%%%%%%%%%%%%%%%%%%%%%%%%%%%%%%%%%%%%%%%%%%%%%%%%%%%%
In this section, we apply the developed methods to the jet engine sensor data. Except for the model introduced in Section~\ref{sec:comparemodel1}, in the real application, we also compare our model with existing models in literature \cite{kim2019generic}, which will be introduced in Section~\ref{sec:real.hi}.

As described in Section~\ref{sec:motivating.example}, the jet engine dataset that we used contains 200 units with 100 failures and 100 censors. To investigate the prediction ability of each model, we split the data into 80\% training set and 20\% testing set. We train the degradation model with the training units and test the parameter estimation with the test units. For each of the comparison methods, we repeat the splitting 50 times and the results are summarized in Section~\ref{sec:real.res}.

%%%%%%%%%%%%%%%%%%%%%%%%%%%%%%%%%%%%%%%%%%%%%%%%
%%%%%%%%%%%%%%%%%%%%%%%%%%%%%%%%%%%%%%%%%%%%%%%%
\subsection{Existing Health Index Model}\label{sec:real.hi}

\citeN{kim2019generic} proposed a latent linear model to construct a health index using multiple sensors and involve variable selection. The health index (i.e., degradation index) is a linear combination of the sensors, as well as a linear combination of basis functions with measurement error, which is denoted as $\hi$ (``KSL'' are the authors' last name initials). For unit $i$, the health index at time $t$ is written as:
\begin{equation*}
h_i(t) = \xvec_i(t) \omegavec_0 = \psivec(t) \tauvec_i + \epsilon_i(t),
\end{equation*}
where $\xvec_i(t)$ denotes the sensor vector, $\omegavec_0$ represents the model parameter vector to be estimated, $\psivec(t)$ are the basis functions, $\tauvec_i$ is the coefficient vector for the basis functions, and $\epsilon_i(t)$ follows a normal distribution. To obtain the estimation of $\omegavec_0$, the property $\psivec(t_i)\tauvec_i = \alpha$ of failed units is used, where $\alpha$ is the failure threshold and $t_i$ is the time to failure for the $i$th failed unit. Therefore, when training the health index model, only the failed units are used. The maximum likelihood method was used to estimate $\omegavec_0$. The variable selection is done by the adaptive LASSO. To ensure a monotonic trend (i.e., decreasing/increasing) of the health index, the paper also introduced strategies to resolve practice issues. In the following model comparison, the choice of basis and tuning parameters of the model is set to be the same as in \citeN{kim2019generic}.

\subsection{Result Comparisons}\label{sec:real.res}

For the model comparison, we randomly split the dataset into 80\% training set and 20\% testing set while keeping the proportion of failed units as 50\% in both training and testing sets. We apply each of the four models $\degmodel$, $\nosel$, $\linear$, and $\hi$ to the training set and then use the trained model to predict unit status in the test set. In $\degmodel$ and $\nosel$ model, we use 10 spline bases to represent each of the 16 sensors, so in total, we have 160 coefficient parameters. The target failure threshold is set as $\alpha = \exp(5)$. For $\hi$, we only use the failed units in the training set to build the model due to the model property. For each of the comparison method, we repeat the splitting 50 times and the results are summarized as follow.

Figure~\ref{fig:real.compare001} presents the boxplot of the prediction errors over 50 splits on the testing set of four models and Table~\ref{tbl:real_fourmodel} provides the average prediction error over repetitions. We can see that $\degmodel$ has the lowest averaged total error and $\FPR$ among the four models. The $\linear$ has zero $\FNR$. However, its $\FPR$ is unusually large. One potential reason is that $\linear$ fails to capture the nonlinear trend in the data can cause the estimation of $\log(\sigma^{\ast})$ fails to shrink as expected, which results in the model having errors on one side. The $\hi$ model has the second smallest average $\FPR$. However, its $\FNR$ is 4\% larger than our proposed model $\degmodel$. One possible reason is that the $\hi$ model neglects the censored unit's information when training the model. Except for $\linear$, the other three models have similar inter quantile range (IQR) for prediction errors.

\begin{figure}
	\centering
	\begin{tabular}{cc}
		\includegraphics[width=0.48\textwidth]{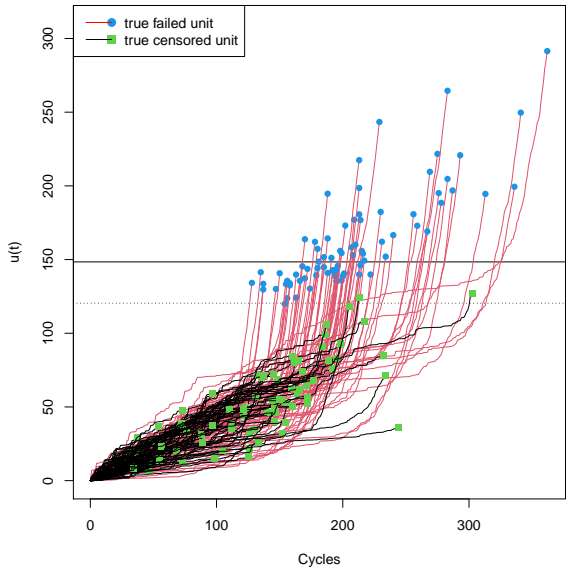}&
		\includegraphics[width=0.48\textwidth]{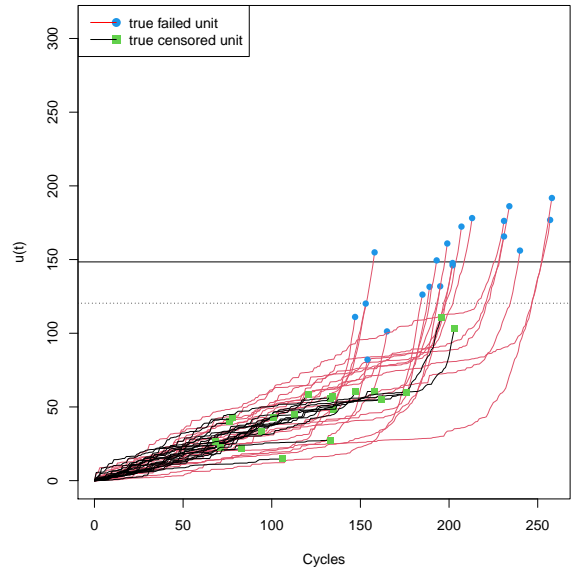}\\
		(a) The $u(t)$ on Training Set & (b) The $u(t)$ on Testing Set\\
	\end{tabular}
	
	\caption{The degradation index for training and testing set with one split of the jet engine data. The horizontal lines represent different thresholds. The solid line is $\alpha = \exp(5)$ and the dotted line is $\widetilde{\alpha}_{0.01}$.}
	\label{fig:realdeg}
\end{figure}

Figure~\ref{fig:realdeg} presents the degradation index build with our proposed framework $\degmodel$ from one split. In this plot, in the testing set, all censored units are below $\widetilde{\alpha}_{0.01}$ threshold and most failed units are over that threshold. We can see from the plot that using a practice failure threshold helps to allow more true failed units' $u(t)$ to reach the threshold.

Regrading to the variable selection, in the above jet engine data analysis, all 16 sensors are kept by the $\degmodel$ model. However, as we have already seen in the simulation study, when the number of units in the model becomes larger, $\degmodel$ tends to keep more variables in the model even some of them are no-effect. Therefore, we take a small subset of the jet engine data and test the model's variable selection ability when $n$ is small. Figure~\ref{fig:ale.var}(a) shows the proportion of each variable in the model excluded over 20 replicates when the number of units changes from 40 to 80 with 10 increment. We can see that when the number of units is relatively small, some sensors such as NRf, altitude, and Nf are excluded from the model with high probability. When $n = 40$, each variable is excluded from the model at least once. However, as $n$ increases, there are variables that can be remained the model for all repetitions.

\subsection{Visualization of Signal Effects on the Degradation Process}

\begin{figure}
	\centering
	\begin{tabular}{cc}
	\includegraphics[width=0.48\textwidth]{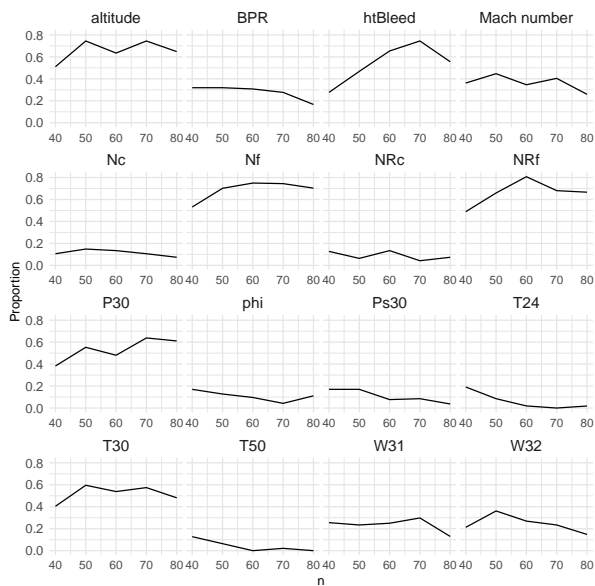}&
    \includegraphics[width=0.48\textwidth]{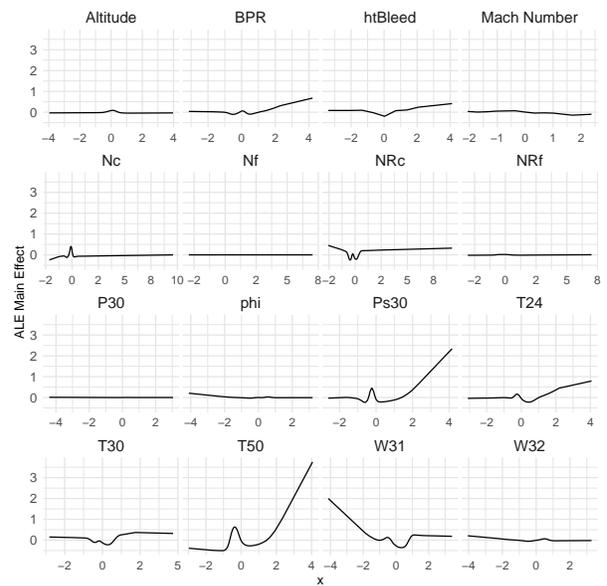}
 \\
(a) Proportion of Sensors Removed & (b) Accumulated Local Effects
	\end{tabular}
	
	\caption{The proportion of sensors be removed from the model when the number of units is small with the jet engine data (a) and accumulated local effects of each signal to the degradation process (b). The $x$-axes in (b) are the standardized measurements.}
	
	\label{fig:ale.var}
\end{figure}

To better understand the sensors effects on the degradation path, we use the accumulated local effects (ALE) plot proposed by \citeN{apley2020visualizing} for visualization. ALE plot is a visualization approach to present the predictors' main and secondary-order effects in complex black box supervised learning models. Here, we focus on the main effect of the sensors. Suppose the model has $p$ predictor variables $\Xvec = (X_1, \dots, X_p)'$ and the $f(\xvec) = f(x_1, \dots, x_p)$ is the fitted model that predicts the response as a function of observed variables $\xvec$. The ALE main effect is defined as
\begin{equation}
g_{j,\ALE}(x_j) = \int_{x_{\min, j}}^{x_j} \mathds{E}[f^j(X_j, \Xvec_{\backslash j})|X_j=z_j] dz_j,
\label{eq:ale}
\end{equation}
\noindent
where $\Xvec_{\backslash j}$ denotes the $p-1$ predictors excluding $X_j$ and $f^j(X_j, \Xvec_{\backslash j}) = {\partial f(\Xvec)}/{\partial X_j}$ represents the local effect of $X_j$ on $f(\Xvec)$. Here we consider visualizing the effect of sensors to the damage exposure for interpretability. That means we use the pre-integration damage level $h\left\{\sum_{i = 1}^p f_j[x_j(s);\betavec_j]\right\}$ in (\ref{eq:u(t)}) instead of $u(t)$ as the predictive model $f(\xvec)$ in (\ref{eq:ale}) to reduce the complexity in interpretation introduced by integration. The ALE plot of each sensor is shown in Figure~\ref{fig:ale.var}(b). The temperature at LPT outlet (T50) and pressure at HPC outlet tend to have constant influence when the measurements are low and larger effect on the damage level when measurements increase. While the coolant bleed (W31) has decreasing effect to the damage level before a certain point then the effect becomes constant. We can also see from Figure~\ref{fig:ale.var}(a) that sensors with no obvious effect (i.e., the ALE main effects are constant as the measurements changing) on the degradation process are more likely removed from the model, such as altitude and NRf.

\begin{table}
\centering
\caption{The average error rate over 50 splits with practical threshold $\widetilde{\alpha}_{0.01}$ of four methods.}
\label{tbl:real_fourmodel}
\vspace{2ex}
\begin{tabular}{l|rrr}
  \hline
  \hline
Method & FNR & FPR & TER\\
  \hline
 DI-VS & 0.030 & 0.026 & 0.057 \\
 DI-KSL & 0.070 & 0.034 & 0.104 \\
 DI-VSL & 0.000 & 0.645 & 0.645 \\
 DI-NVS & 0.040 & 0.040 & 0.080 \\
   \hline
   \hline
\end{tabular}
\end{table}

%%%%%%%%%%%%%%%%%%%%%%%%%%%%%%%%%%%%%%%%%%%%%%%%%%%%%%%%%%%%%%%%%%%%%%%%%%%%%%%%%%%%%%%%%%%%%%%%%%%%%%%%%%%%%%%%%
\section{Conclusions and Areas for Future Research}\label{sec:conclusion}

In this paper, motivated by the jet engine multi-channel sensory data, we propose a new framework to build the degradation index based on the cumulative exposure model. The framework is flexible and we adopt the M-spline basis to model arbitrary functional forms of degradation signals from sensors. The nonlinear transformation $h(z)$ in the proposed model ensures the monotonicity of the degradation index. To obtain the parameter estimation, the log-likelihood is used as the main component in the objective function to handle both failed and censored units. The adaptive group LASSO is also involved to perform variable selection in the model.

We demonstrate the proposed model's performance in two aspects, one is the ability to correctly predict a product status, the other is variable selection. We conduct comprehensive simulation studies and perform jet engine data analysis. We compare our proposed model with the model that only assumes linear effect of sensors, the model without variable selection, and the existing health index model.

For the prediction accuracy, because we focus on the $\FNR$ in our application, we propose to use an asymmetric LEV distribution and a practice threshold to avoid falsely predict a failed product as censored. The study shows that our proposed framework performance is more robust than other models. $\degmodel$ has consistently good prediction accuracy regardless of the dataset size and scenarios. Although $\nosel$ also has good prediction accuracy when the $n$ in the dataset is large, it has a large $\FNR$ when $n$ is small. This illustrates the benefit of involving variable selection in the model especially when $n$ is small. Including all sensors in the model can help to fit a good model when the sample size is small but does not contribute when making predictions. $\linear$ model also has good prediction ability in some scenarios. However, it can have a very large $\FPR$ in certain cases. This shows that failing to capture the nonlinearity in the data may lead to extreme model performance. Compared to the $\hi$ model in the real application, our proposed model outperforms in both $\FNR$ and $\FPR$. One possible reason is that we make use of both censored and failed products information when training the model. Besides, unlike mentioned in \citeN{kim2019generic}, we do not need to manually pre-screen sensors or impose constraints to coefficient parameters to ensure the trend of the degradation index. Our model automatically guarantees the monotonicity of the degradation index.

Regarding the variable selection ability in the proposed approach, our study results show that as the number of units $n$ in the dataset increases, the model is able to keep more correct sensors but harder to exclude no-effect sensors. One possible reason is that we use the log-likelihood function in the objective function. As the number of units increases, as long as the sensor can provide a little contribution to the likelihood, adding up these contributions of $n$ units may lead to large minimization of the objective function. So when $n$ is large, it is hard to exclude no-effect sensors from the model.

In this paper, we mainly focus on using the degradation index to predict product status. However, the usage of this degradation index is not limited to the status prediction. Existing models of degradation index can be used for the product health assessment. Besides the potential usage of the proposed model, we also want to discuss the flexibility of this framework. Although in this paper, we present an additive nonlinear manner to model the sensors' contribution and use LEV distribution to construct the objective function, there are much more ways to use this framework. People can change how to model the impact of sensors $f_j\left(x_j\right)$, whether to allow interactions between sensors effect, and how to conduct the nonlinear transformation $h(z)$ based on their knowledge about different applications. Besides, if in an application, other criteria such as $\FPR$, $\TER$ are focused, the choice of distribution in the objective function can also be changed correspondingly. Thus, what we provide in this paper can be considered as a general flexible framework rather than a fixed model targeting a specific problem.

There are some future directions for the proposed model. In this paper, we consider an additive way to model all sensors' impact. However, in reality, the way that sensors influence the degradation process can be much more complex. It is interesting to allow sensors interactions in the model.
Besides, it would be useful to improve the model's ability to maintain the useful variables and exclude no effect sensors. A penalty term in the adaptive group elastic net manner can be considered. Also, it is interesting to involve the number of units in the penalty when conducting the variable selection.

%%%%%%%%%%%%%%%%%%%%%%%%%%%%%%%%%%%%%%%%%%%%%%%%%%%%%%%%%%%%%%%%%%%%%%%%%%%%%%%%%%%%%%%%%%
\section*{Acknowledgment}

The authors acknowledge the Advanced Research Computing program at Virginia
Tech for providing computational resources. The work by Hong was partially
supported by National Science Foundation Grant CMMI-1904165
to Virginia Tech.

%%%%%%%%%%%%%%%%%%%%%%%%%%%%%%%%%%%%%%%%%%%%%%%%%%%%%%%%%%%%%%%%%%%%%%%%%%%%%%%%%%%%%%%%%%%%%%%%%
%\bibliographystyle{chicago}
%\bibliography{ref}

\end{document}